\documentclass[twocolumn]{revtex4}
\usepackage{float}
\usepackage{psfrag,color,pst-grad}
\usepackage{pstricks}
\usepackage{epsfig}
\usepackage{subfigure}
\usepackage{graphicx}
\usepackage{graphics}
\usepackage[latin1]{inputenc}
\usepackage[T1]{fontenc}
\usepackage{times}
\usepackage{amsfonts}
\usepackage{amssymb}
\usepackage{amsmath}
\usepackage{latexsym}

\newcommand{\be}{\begin{equation}}
\newcommand{\ee}{\end{equation}}
\newcommand{\beqn}{\begin{eqnarray}}
\newcommand{\eeqn}{\end{eqnarray}}
\newcommand{\rd}{{\rm d}}

\newcommand{\ti}{\textit}
\newcommand{\Frac}[2]{\frac{{\displaystyle #1}}{{\displaystyle #2}}}



\begin{document}

\begin{abstract}

An analysis of the solutions for the field equations of generalized
scalar-tensor theories of gravitation is performed through the study
of the geometry of the phase space and the stability of the solutions,
with special interest in the Brans-Dicke model. Particularly, we
believe to be possible to find suitable forms of the Brans-Dicke
parameter $\omega$ and potential V of the scalar field, using the
dynamical systems approach, in such a way that they can be fitted in
the present observed scenario of the Universe.
\end{abstract}

\title{Phase space solutions in scalar-tensor cosmological models}
\author{Jos\'e C. C. de Souza}
\affiliation{Departamento de F\'isica-Matem\'atica, IFUSP}
\author{Alberto Saa}
 \affiliation{IMECC-Unicamp}
\maketitle

\section{Scalar-Tensor Theories of Gravitation} 

 In a homogeneous and isotropic space, described by the
 Friedmann-Lemaître-Robertson-Walker metric
\be\label{metric}
\rd s^2=-\rd t^2 + a^2(t)\left[\Frac{\rd r^2}{1- K r^2}+ r^2(\rd
  \theta^2+\sin^2\theta \rd \varphi^2)\right],
\ee
\noindent where $a$ is the scale factor and $K$ is the spatial
curvature index, gravitation can be described by an action of the kind
\be\label{action} \small{
S = \Frac{1}{16 \pi} \int \rd^4 \sqrt{-g}\left[\phi R
  -\Frac{\omega(\phi)}{\phi} g^{ab}\nabla_{a} \phi\nabla_{b}
  \phi-V(\phi)\right] + S^{m}} ,
\ee 
\noindent where $S^{m}$ is the action of usual matter, $g$ is the
determinant of the metric tensor, $\omega$ is a coupling function
(which we will eventually assume to be a constant, known as
Brans-Dicke parameter) and $V(\phi)$ is the potential of the scalar
field $\phi$ \cite{Faraoni}.

From (\ref{action}), we obtain for the field equations:
\be\label{H_quad}
H^2 = -H\left(\Frac{\dot\phi}{\phi}\right) +
\Frac{\omega}{6}\left(\Frac{\dot\phi}{\phi}\right)^{2} +
\Frac{V(\phi)}{6\phi}-\Frac{K}{a^2} +\Frac{8\pi\rho^{m}}{3\phi},
\ee
\beqn
\dot{H}&=&
-\Frac{\omega}{2}\left(\Frac{\dot\phi}{\phi}\right)^{2}+2 H
\left(\Frac{\dot\phi}{\phi}\right) \nonumber \\&&  +
\Frac{1}{2(2\omega+3)\phi}\left[\phi\Frac{\rd V}{\rd \phi}-2 V +
  \Frac{\rd \omega}{\rd \phi} (\dot\phi)^{2}\right] \nonumber \\&& +
\Frac{K}{a^{2}}-\Frac{8\pi}{(2\omega+3)\phi}[(\omega+2)\rho^{m}+ \omega
  P^{m}],
\eeqn
\beqn 
\ddot\phi+\left(3H+\Frac{1}{2\omega+3}\Frac{\rd \omega}{\rd
\phi}\right)\dot\phi &=& \Frac{1}{2\omega+3} \left[2V -\phi\Frac{\rd
    V}{\rd \phi} + \right.\nonumber \\ &&  \left.8\pi(\rho^{m}-3P^{m}) \right] ,
\eeqn
\noindent where $H\equiv \dot{a}/a$ is the Hubble parameter and
$\rho^{m}$ and $P^{m}$ are the energy density and the pressure of the
material fluid.

As usual, we parameterize the equation of state for the fluid as
$P^{m} =(\gamma -1)\rho^{m}$ with $\gamma$ a constant chosen to
indicate a variety of fluids that are predominantly responsible for
the energy density of the Universe. We can see that through the energy
conservation equation $\dot\rho^{m} + 3H(\rho^{m}+ P^{m} )= 0$ we
obtain $\rho^{m}=\rho_{0}/{a^{3\gamma}}$, with $\rho_{0}$ a
constant.

\section{The case for $V=\frac{1}{2}m^2\phi^2$ and $K=0$}

In the referred paper \cite{Faraoni}, the author proceeds to show the
phase space allowed for the orbits of solutions for these field
equations in several cases with different potentials and parameters
$\omega$. For example, in the case of vacuum, flat space ($K=0$) and
potential $V=\frac{1}{2}m^2\phi^2$, equation (\ref{H_quad}) was
rearranged as (making $m\equiv 1$)
\be
\omega\dot{\phi}^2-6H\phi\dot{\phi}+(\Frac{1}{2}\phi^2-6H^2\phi)\phi=0,
\ee
\noindent which has the solutions
\be\label{phi_dot}
\dot{\phi}_{\pm}(H,\phi)=
\Frac{1}{\omega}\left[3H\phi\pm\sqrt{3(2\omega+3)H^2\phi^2-\frac{1}{2}\omega\phi^2}\right].
\ee
The assumption of flat space is required in order to reduce the
dimensionality of the phase space.

We want to analyze qualitatively the geometry of the phase space
$(H,\phi,\dot\phi)$, expecting to infer the form of the functions
$\omega(\phi)$ and $V(\phi)$ to fit better the available data on the
structure of the Universe.

The phase space for this situation is composed of a 2-d surface with
two sheets, related to the lower and upper signs in
eq. (\ref{phi_dot}). Figures 1-3 show the phase space for the choice
$\omega=10$.

\begin{figure}
\includegraphics[scale=0.4]{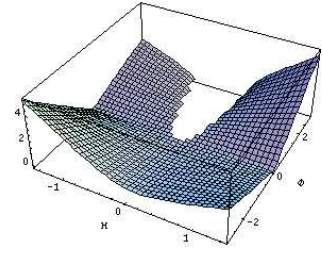}
\caption{Upper sheet of the of the phase space for a model with $\omega=10$
(Brans-Dicke), corresponding to the positive sign in
eq. (\ref{phi_dot}) \cite{Faraoni}. The ``hole'' in the surface
indicates the region forbidden for the orbits of the solutions.}
\end{figure}

\begin{figure}
\includegraphics[scale=0.4]{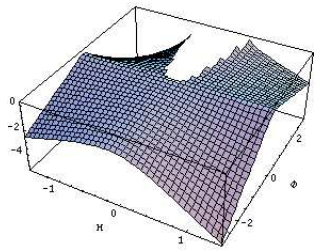}
\caption{Lower sheet of the phase space, now corresponding to the negative sign in
eq. (\ref{phi_dot}).}
\end{figure}

\begin{figure}
\includegraphics[scale=0.4]{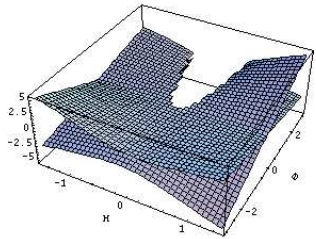}
 \caption{The complete phase space composed of the upper and lower sheets
  linked to each other at the boundary of the forbidden region.}
\end{figure}

 The fixed points for this dynamical system, obtained making
 $\dot{H}=\dot{\phi}=0$, are de Sitter solutions, given by $H_{0}=\pm
 \sqrt{{\phi_{0}}/{12}}$, with constant $H_{0}$ and $\phi_{0}$.

\section{The case for $V=\Lambda\phi$ and $K=0$}

Following other works (\cite{Kolitch}-\cite{Santos}) which give a
complete analysis of the phase space for Brans-Dicke model with a
cosmological constant $\Lambda$ (simply making $V(\phi)=\Lambda \phi$
in the action), we can illustrate the situation in which $\omega$ has
a very large value and $\gamma=0$. Therefore, the energy density of
the fluid is a constant $\rho_{0}$. It should be emphasized that
recent observational and simulation results seem to favor a scenario
very similar to this one
(\cite{Sanchez},\cite{Esposito2}-\cite{Liddle}). The solutions in this
case are written as \be\label{phi_dot2}
\small{\dot{\phi}_{\pm}(H,\phi)=
\Frac{1}{\omega}\bigg[3H\phi\pm\sqrt{9H^2\phi^2-\omega[\phi^2(\Lambda-6H^2)+16\pi\rho_{0}]}\bigg]}.
\ee

With this solutions, we are able to show the phase space for a
particular choice of constants $\Lambda$, $\omega$ and $\rho_{0}$.

We proceed to find the dynamical equations system for this simple
model, as done before.

\begin{figure}
\includegraphics[scale=0.4]{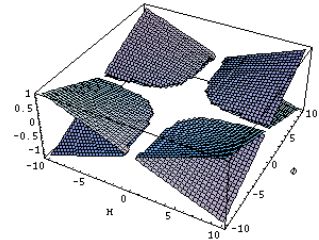}
\caption{Complete phase space for a Brans-Dicke model with a
cosmological constant $\Lambda=1$, energy density $\rho_0=2$ and
constant parameter $\omega=50000$, showing two sheets linked by the
boundary of the forbidden region, as in the precedent case.}
\end{figure}

Naming $\Delta$ the expression under the root in eq.(\ref{phi_dot2}),
we can write the equation for $\dot{H}$: \beqn\label{dot_H}
\dot{H}_{\pm}&=&
-\Frac{1}{2\omega\phi^2}\left[3H\phi\pm\sqrt{\Delta}\right]^2 +
\Frac{2H}{\omega}\left[3H\pm\Frac{\sqrt{\Delta}}{\phi}\right]\nonumber
\\&& -\Frac{1}{2(2\omega+3)}\left(\Frac{\Lambda}{2}
+\Frac{16\pi\rho_{0}}{\phi}\right).  \eeqn Now, equations
(\ref{phi_dot2}) and (\ref{dot_H}) form the system for which the fixed
points are the solutions $H_{0}=\pm
\sqrt{{8\pi\rho_{0}}/{3\phi_{0}^2}+{\Lambda}/{6}}$ .

Of special interest is the search for the most adequate functions
$\omega(\phi)$ and $V(\phi)$, that may be more complicated than what
was assumed until here.

\section{Conclusions}

The method of analyzing the geometry of the phase space have proved to
be a useful tool in the search for the solutions of the field
equations of generalized gravity models. Our aim is to achieve a
complete analysis of the simple model presented before (including the
stability of the solutions, via Lyapunov's direct method
\cite{LaSalle}, in order to investigate further its \textit{attraction
basin}) and to apply more sophisticated functions to it.

\section{Acknowledgments}

This work is supported by Conselho Nacional de Pesquisa e
Desenvolvimento (CNPq).


\begin{thebibliography}{99}

\bibitem{LaSalle} J. LaSalle and S. Lefschetz, ``Stability by
  Lyapunov's Direct Method''. Academic Press (1967)

 \bibitem{Faraoni} V. Faraoni, \ti{Annals of Physics} {\bf 317} (2005)
 366

 \bibitem{Kolitch} S. J. Kolitch, \ti{Annals of Physics} {\bf 246}
 (1996) 121

 \bibitem{Kolitch2}S. J. Kolitch, \ti{Annals of Physics} {\bf 241}
 (1995) 128

\bibitem{Santos}C. Santos and R. Gregory, \ti{Annals of Physics} {\bf
  258} (1997) 111

 \bibitem{Sanchez} A. G. Sanchez et al., astro-ph/0507583

 \bibitem{Esposito} G. Esposito-Farèse and D. Polarski,
  \textit{Phys. Rev.} {\bf D63} (2001) 063504

 \bibitem{brussels} A. Saa et al., \textit{Phys. Rev.} {\bf D63} 
 067301 (2001); \textit{Int. J. Theor. Phys. } {\bf 40}, 2295 (2001);
 L.R. Abramo, L. Brenig, E. Gunzig, A. Saa, \textit{Phys. Rev.} {\bf
 D67} (2003) 027301; gr-qc/0305008.

 \bibitem{Barrow} J. D. Barrow and J. P. Mimoso, \textit{Phys. Rev.}
 {\bf D50(6)} (1994) 3746

 \bibitem{Fabio} F. C. Carvalho and A. Saa, \textit{Phys. Rev.} {\bf
 D70} (2004) 087302

 \bibitem{Esposito2} G. Esposito-Farèse, gr-qc/0409081

 \bibitem{cassini} B. Bertotti, L. Iess and P. Tortora,
 \textit{Nature} {\bf 425} (2003) 374-376
 
 \bibitem{Acquaviva} V. Acquaviva, C. Baccigalupi, S. M. Leach, Andrew
 R. Liddle and F. Perrotta, astro-ph/0412052

 \bibitem{Liddle} A. R. Liddle, A. Mazumdar and J. D. Barrow,
 \textit{Phys. Rev.} {\bf D58} (1998) 027302

  \end{thebibliography}
\end{document}